\documentclass{aa}

\usepackage{latexsym}
\usepackage{graphicx}
\usepackage{natbib}
\usepackage[latin1]{inputenc}
\usepackage{color}

\newcommand{\sect}[1]{Sect.\,\ref{#1}}
\newcommand{\fig}[1]{Fig.\,\ref{#1}}

\definecolor{dgreen}{rgb}{0,0.5,0}

\begin{document}

\title{Ejection of cool plasma into the hot corona}

\author{P. Zacharias \inst{1,}\thanks{\emph{Present address:} International Space Science Institute, Hallerstrasse 6, CH-3012 Bern, Switzerland} \and H. Peter \inst{2}\and S. Bingert\inst{2}}
\institute{Kiepenheuer-Institut f\"ur Sonnenphysik, Sch\"oneckstrasse 6,
  79104 Freiburg, Germany \and 
Max Planck Institute for Solar System Research, Max-Planck Strasse 2, 37191 Katlenburg-Lindau, Germany}

\abstract%
{
The corona is highly dynamic and shows transient events on various scales in space and time. Most of these features are related to changes in the magnetic field structure or impulsive heating caused by the conversion of magnetic to thermal energy.
}{
We investigate the processes that lead to the formation, ejection and fall of a confined plasma ejection that was observed in a numerical experiment of the solar corona. By quantifying physical parameters such as mass, velocity, and orientation of the plasma ejection relative to the magnetic field, we provide a description of the nature of this particular plasma ejection.
}{
The time-dependent three-dimensional magnetohydrodynamic (3D MHD) equations are solved in a box extending from the chromosphere, which serves as a reservoir for mass and energy, to the lower corona. The plasma is heated by currents that are induced through field line braiding as a consequence of photospheric motions included in the model. Spectra of optically thin emission lines in the extreme ultraviolet range are synthesized, and magnetic field lines are traced over time. We determine the trajectory of the plasma ejection and identify anomalies in the profiles of the plasma parameters.
}{
Following strong heating just above the chromosphere, the pressure rapidly increases, leading to a hydrodynamic explosion above the upper chromosphere in the low transition region. The explosion drives the plasma, which needs to follow the magnetic field lines. The ejection is then moving more or less ballistically along the loop-like field lines and eventually drops down onto the surface of the Sun. The speed of the ejection is in the range of the sound speed, well below the Alfv\'en velocity. 
}{
%
The plasma ejection observed in a numerical experiment of the solar corona is basically a hydrodynamic phenomenon, whereas the rise of the heating rate is of magnetic nature. The granular motions in the photosphere lead (by chance) to a strong braiding of the magnetic field lines at the location of the explosion that in turn is causing strong currents which are dissipated.  Future studies need to determine if this process is a ubiquitous phenomenon on the Sun on small scales. Data from the Atmospheric Imaging Assembly on the Solar Dynamics Observatory (AIA/SDO) might provide the relevant information.
}

\maketitle

\section{Introduction}

The solar corona is highly structured and very dynamic. This ranges from large disruptive events such as flares and coronal mass ejections (CMEs) down to small-scale explosive events in the transition region between the chromosphere and the corona. Because the coronal dynamics are a direct signature of the processes governing the energy balance in the upper atmosphere, i.e., coronal heating, radiative losses and heat conduction, the investigation of coronal transients on all temporal and spatial scales is a key for also understanding the nature of the corona of the Sun and other stars. We present results from a three-dimensional magnetohydrodynamic (3D MHD) numerical experiment of the solar corona in which an explosion-like event drives a confined plasma ejection ballistically through the corona along the magnetic field lines.

Imaging instruments provide information on the temporal and spatial variation of the coronal emission. Especially during transient events, their interpretation is often non-trivial and might be attributed to either waves or mass flows. For example, early on,  intensity variations were associated with the evaporation and condensation of plasma and the motion of the condensation region \citep{antiochos+sturrock:1978}. Outward-propagating features of fan-like structures have often been interpreted as (slow) magneto-acoustic waves, \citep[e.g.,][]{demoortel:2002a}.

The investigation of cool plasma of several 10.000\,K as seen in the Ly-$\alpha$ line above the limb indicates that small brightenings are moving along the magnetic field lines. These have been analyzed in depth by \cite{Schrijver:2001} using data from TRACE \citep[Transition Region and Coronal Explorer;][]{1999SoPh..187..229H}. The movie provided with the online version of \cite{Schrijver:2001} clearly shows three types of bright cool structures moving in the corona: (1) features appearing up in the corona that are falling down, (2) brightenings moving upward that subsequently disappear, and (3) small brightenings initially moving upward before they fall back again. These types of brightenings have been interpreted in different ways. 

Type (1) features can be interpreted as condensations in the corona caused by a loss of thermal equilibrium. These features have been investigated in depth by \cite{mueller:2003,mueller:2004} through 1D loop models and compared to observations \citep{mueller:2005}. These models show that if the heating is concentrated in the lower atmosphere and is not sufficient to balance the radiative cooling at the loop apex, a runaway process will be initiated, because the cooler plasma can cool more efficiently and through this rapidly forms a condensation. Eventually, this condensation slides down the field lines with speeds of up to 100\,km/s. \cite{degroof:2004} observed that features moving downward are indeed plasma flows rather than magneto-acoustic waves.

The second type of cool brightenings that move upward and subsequently disappear might be associated with the high-speed transition region line upflows as discussed by \cite{mcintosh+depontieu:2009}. The authors conclude that these upflows might be visible as propagating disturbances in the extreme ultraviolet. They may or may not be the small-scale relatives of explosive events and spicules. \cite{mcintosh+depontieu:2009} and \cite{depontieu:2009} found these upflows through a blueward asymmetry and line broadening of transition region and coronal lines in active regions that have been reported before for transition lines in the quiet Sun network \citep{peter:2000,peter:2001}. The cool brightenings eventually heat up and disappear from the observations of the cool plasma (as in Ly-$\alpha$), leaving a footprint in the asymmetry of lines formed at higher coronal temperatures above $10^6$\,K \citep{hara+al:2008,peter:2010}.

The type (3) cool brightenings that move up and fall back might be interpreted as plasma that is propelled upward, follows the magnetic field lines and then falls back onto the Sun without being heated up to coronal temperatures as the type (2) brightenings are.
We found evidence for this type (3) process in our numerical experiments of the solar corona. The driving of the magnetic field in the photosphere through horizontal convective (granular) motions results in a high energy input over a short time caused by Ohmic dissipation in the transition region. An explosion event of this kind would be pushing cool material upward, which then follows the magnetic field lines up to the apex and slides down on the other side of the magnetic loop.

The high energy deposition that eventually leads to the plasma ejection comes about self-consistently when (by chance) the granular motions in the photosphere lead to a magnetic configuration with a high rate of Ohmic dissipation in a small region in the simulation box.

By analyzing the data of the numerical experiment, we will investigate the formation, rise and fall of the plasma ejection in the 3D MHD model and quantify its physical parameters, such as mass, velocity and orientation relative to the magnetic field. In addition, we synthesize emission lines that are observable with current instrumentation.

We emphasize that this study is not investigating spicule-like phenomena. Related high-speed upflows that have recently been observed from the chromosphere through the transition region to the corona \citep{depontieu+al:2011} most likely play a significant role for the supply of hot plasma into the corona. 
Likewise, observations have shown strong upflows at the edges of active regions \citep[e.g.,][]{sakao+al:2007,Doschek+al:2008}. Observational evidence exists that these upflows appear as asymmetries in the line profiles \citep[e.g.,][]{depontieu:2009,mcintosh+depontieu:2009} or as second components in the spectral profiles \citep[e.g.,][]{hara+al:2008,peter:2010}.
However, these high-speed upflow structures seem to be relatively stable.
Probably, the transient plasma ejection we discuss here is closer related to miniature coronal mass ejections (mini-CMEs) as described by, e.g., \cite{Innes+al:2009,Innes+al:2010}.

We briefly explain the setup of the numerical expriment in \sect{model}.
A qualitative description of the observed phenomenon is provided in Sect.\,\ref{phenomenology}, and the onset of the eruption is investigated in detail in Sect.\,\ref{onset}. The dynamic evolution of the ejection along its trajectory is discussed in Sect.\,\ref{evolution}, and in Sect.\,\ref{aia_section} we comment on the visibility of this phenomenon with respect to the Atmospheric Imaging Assembly onboard the Solar Dynamics Observatory (AIA/SDO). Finally, the results are summarized and discussed in Sect.\,\ref{discussion}.

\begin{figure*}
\includegraphics[width=\textwidth]{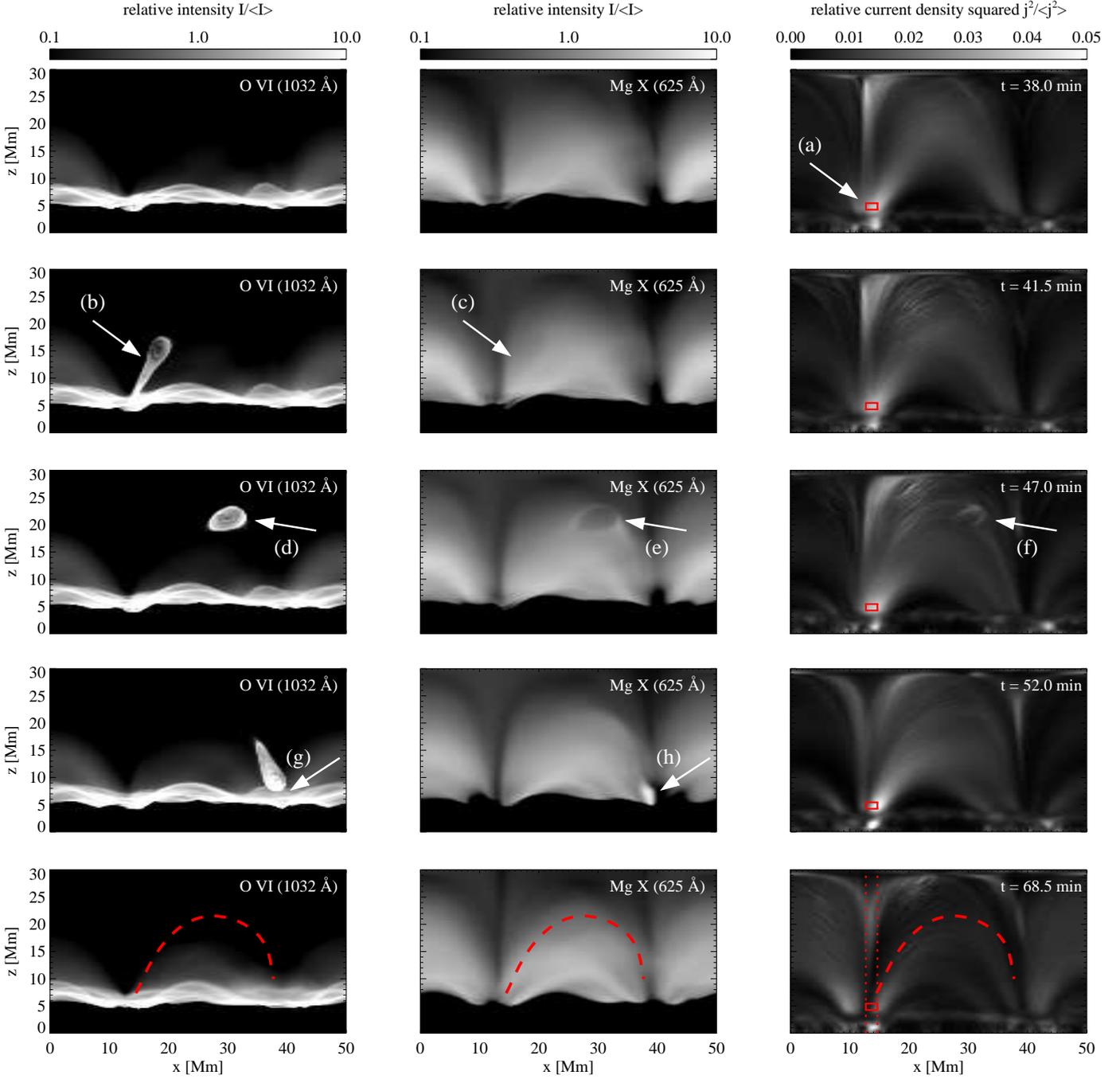}
\caption{Evolution of the plasma ejection over 30 minutes (top to bottom, times noted in right panels).
The left and middle panels show the (normalized) synthesized emission in O\,{\sc{vi}} (1032\,\AA) and Mg\,{\sc{x}} (625\,\AA) integrated horizontally through the computational box. This corresponds to what would be observed at the solar limb.
The right panels show the currents, $j^2$, i.e., the Ohmic heating rate, also integrated horizontally through the box. $j^2$ is divided by the average trend with height to account for the roughly exponential decrease of the heating rate with height \citep[cf.][]{bingert:2010}.
The arrows point to the same location in the maps in each row.
The red line in the bottom row indicates the trajectory of the plasma eruption that was deduced from the position of its center-of-gravity (see section \ref{formation2}). The red box in the $j^2$-images shows the approximate formation region of the plasma eruption; the dotted lines indicate its extension in the $x$-direction.
See also movie in Appendix \ref{appendix}, Fig.\,\ref{fig.movie}.
}     
\label{emiss}
\end{figure*}

\section{Setup of the numerical experiment}\label{model}

We analyze an ejection of cool material into the corona in a 3D MHD simulation of the solar corona above a small active region.
The setup of the 3D model is identical to the one described in \cite{bingert:2010}. We used the Pencil code \citep{brandenburg+dobler:2002} to solve the 3D MHD equations. The heat conduction term parallel to the magnetic field and the radiative losses were included in the energy equation to obtain a realistic description of the coronal pressure and thus of the emerging synthesized emission from the transition region and corona. The corona is heated through Ohmic dissipation of currents that are induced by the photospheric footpoint motions shuffling around the magnetic field. The computational domain encompasses 50\,Mm ${\times}$ 50\,Mm horizontally and 30\,Mm vertically. The initial magnetic field at the lower boundary is taken from an observed active region magnetogram. After approximately 30\,min of simulation time, the solution is independent of the initial condition, and the magnetic field in the corona is computed self-consistently based on the MHD equations. After this time period, the corona has reached a dynamic equilibrium, where the overall appearance and spatially averaged quantities do not change significantly, while the situation is still dynamic with strong flows and changes in the plasma properties (cf.\ movie attached to \fig{fig.movie}). For more details, we refer the reader to \cite{bingert:2010}.

The model does not contain the proper physics of the chromosphere, especially (non-LTE) radiative transfer is not included in the calculations. However, because the phenomenon we describe emerges from \emph{above} the chromosphere, this is not expected to be a severe issue. At the top of the chromosphere, the energy input is typically $10^{-5}\,\mbox{Wm}^{-2}$, while for the plasma ejection described here, the heating rate just \emph{above} the chromosphere rises to values a factor of 100 higher than that just before the ejection. Therefore, we do not expect the inclusion of non-LTE radiative transfer to have a major effect here. Additional energy transfer, e.g., through radiation, would certainly slightly mitigate the effect, but would not fully compensate the strong increase in heating rate (see also \sect{energy}).

From the results of the 3D MHD model, we calculated the emission in the transition region and corona.
Emission line spectra of various transition region and coronal lines are synthesized using the CHIANTI atomic data base \citep{dere+al:1997,landi+al:2006}. This procedure follows \cite{zacharias:2009,zacharias:2010.doppler}. The O\,{\sc{vi}} (1032\,\AA) line was chosen to represent cool transition region material, and the Mg\,{\sc{x}} (625\,\AA) line to represent hot coronal plasma (see Table\,\ref{lines}). The emission of these lines is roughly comparable to the 304\,\AA\ and 171\,\AA\ bands observed with AIA/SDO (cf.\ Sect.\,\ref{aia_section}), which were derived using the temperature response functions \citep{Boerner+al:2011} as provided by the instrument team in SolarSoft.%
\footnote{http://www.lmsal.com/solarsoft/}
The plasma is assumed to be optically thin and in a state of ionization equilibrium (Peter et al., \citeyear{peter+al:2006}). The constant elemental abundances are taken from Mazotta et al. (\citeyear{mazzotta+al:1998}). 

The specific choice of the O\,{\sc{vi}} and Mg\,{\sc{x}} lines is not crucial. The 
temporal evolution of the ejection looks quite similar in C\,{\sc{iv}} and Ne\,{\sc{viii}}, respectively. The synthesized Mg\,{\sc{x}} images are similar to those synthesized for the 195\,{\AA} channel of AIA, the Ne\,{\sc{viii}} images are similar to those of the 171\,{\AA} AIA band. In this study, the difference between these four coronal images is not significant, and therefore the validity of the images displayed in Figs.\,\ref{emiss}, \ref{aia} and \ref{fig.movie} does not depend on the choice of the respective line. Also, the small differences in appearance between images of C\,{\sc{iv}}, O\,{\sc{vi}} and AIA\,304\,{\AA} are not significant for the illustrative purpose of these figures. Actually, O\,{\sc{vi}} is more similar to the AIA\,304\,{\AA} band than C\,{\sc{iv}}. The AIA\,304\,{\AA} band has some contribution from high temperatures, which originates from hot emission lines in the bandpass. Likewise, O\,{\sc{vi}} shows some contribution from high temperatures that is caused by di-electronic recombination. C\,{\sc{iv}}, being a Li-like ion, also shows this effect, but only to some extent.
The quantitative analysis in Sects.\,\ref{onset} and \ref{evolution} is not influenced by the choice of the lines or bands, of course.

\begin{table}
\begin{center}
\caption{Emission lines and AIA/SDO bands synthesized from 3D MHD model.\label{lines}}
\begin{tabular}{lrc}
\hline
\hline
line & wavelength & formation temperature\\
     & [\AA]        & $\log ( T_{\rm } / [\mbox{K}] ) $ \\
\hline
O\,{\sc{vi}}   &  1032 & 5.5\\
Mg\,{\sc{x}}   &   625 & 6.0\\
\hline
AIA / He~{\sc{i}} &   304 & 4.9 \\
AIA / Fe~{\sc{ix/x}} & 171 & 5.9\\
\hline
\end{tabular}
\end{center}
{\itshape Note 1:} For O\,{\sc{vi}} and Mg\,{\sc{x}}, the line formation temperature is given, i.e., the temperature where 
$G(T)\propto(n_{\rm ion}/n_{\rm el})~T^{-1/2}\,\exp{\big[{-}h\nu / (k_B T)\big]}$ 
reaches its maximum value (assuming ionization equilibrium).
\\
{\itshape Note 2:} For the AIA bands, the main contributing lines are given along with the center wavelength and the formation temperature of the plasma dominating the emission in the respective band.
\end{table}

\section{Phenomenology}\label{phenomenology}

We are focusing on a time sequence of about 70 minutes duration in the numerical experiment. All points of time discussed here refer to time $t{=}0$ at the beginning of this sequence. The actual simulation was started long before $t{=}0$, so that any impact of the initial condition is excluded during the investigated time sequence.

The ejection lifts off at about $t{=}40$\,min and lasts for about 11\,min. It is observed in the form of a ``bubble'' of plasma that rises and falls back down again along a curved trajectory. 
Snapshots of the emissivities integrated along the horizontal $y$-direction of the simulation box are presented in Fig.\,\ref{emiss} for different time steps of the simulation run. In addition, the corresponding Ohmic heating rates ($\propto j^2$, the current density squared)
are shown. For details on the displayed images and a movie of the temporal evolution, see Appendix \ref{appendix} and Eqs.\,\ref{E:jnorm} and \ref{E:jfluct}).

Before the plasma eruption, an increased heating rate is observed just above the chromosphere in the region outlined by the red box and arrow $a$ in Fig.\,\ref{emiss} at $t{=}38$\,min (best seen in the movie with Fig.\,\ref{fig.movie}, indicated by an arrow). At $t{=}41.5$\,min, the plasma ejection lifts off (Fig. \ref{emiss}, second row), and the cool ejection is clearly visible in the O\,{\sc{vi}} line formed at transition region temperatures of about $3{\cdot}10^5$ K (arrow $b$). The cool ejection, however, is hardly visible in the Mg\,{\sc{x}} coronal line formed at approximately $10^6$ K (arrow $c$). In this line, the plasma ejection merely appears as a darkish shade because of its low temperature compared to the surroundings. At $t{=}47$\,min, the ejection reaches its maximum height and starts descending. When falling down, the ejection compresses the magnetic field at its front. As a consequence, currents increase in front of the ejection (arrow $f$), plasma is heated, and the front is brightening (arrows $d,e$; again, best visible in the movie). The magnetic field compression caused by the ejection produces a heating front that is traveling ahead of the actual plasma ejection. Thus, the plasma is heated at low heights just above the chromosphere and consequently shines bright in the coronal line (arrow $h$). At this location and time ($t{=}52$\,min),  the emission in the transition region line is not yet brightening (arrow $g$), but it does when the ejection actually hits the chromosphere (as can be seen in the movie). There is a time delay between the brightening of the coronal line and the transition region line (at the location of arrows $g$ and $h$) of about 1\,min. In the movie, some of the plasma can also be seen to bounce up again in the rear part of the ejection because of the increased pressure when the ejection hits the chromosphere. This observation is similar to the bouncing of condensations falling down on the solar surface, as was found by \cite{mueller:2003,mueller:2004}.
A few minutes later, the plasma ejection has completely disappeared (Fig.\,\ref{emiss}, bottom row).

\begin{figure}
\begin{center}
\includegraphics[width=1.0\columnwidth]{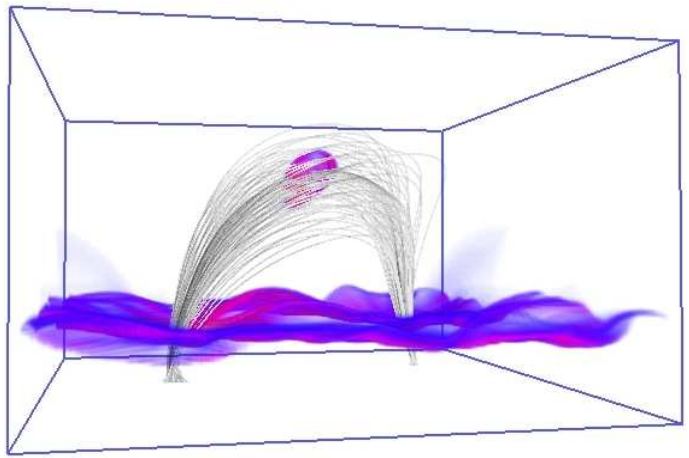}
\caption{Coronal box as seen in O\,{\sc{vi}} (1032 \AA) synthesized from 3D MHD model (emission increases from blue over violet to red). Overplotted are the magnetic field lines that cover the trajectory of the ejection (snapshot at 45\,min).
Visualization using VAPOR \citep{clyne+rast:2005,clyne+al:2007}. 
}
\label{vapor}
\end{center}
\end{figure}

The correlation of the plasma ejection and the magnetic field is depicted in Fig.\,\ref{vapor}, which provides a 3D view of the computational domain as it would be seen in the emission of the O\,{\sc{vi}} line at time $t{=}45$\,min, i.e., close (in time) to the middle row of Fig.\,\ref{emiss}. 

The ejection, shown together with magnetic field lines in the vicinity of its trajectory in Fig.\,\ref{vapor}, has almost reached its maximum altitude and is moving nearly parallel to the magnetic field structure (cf.\,Sect.\,\ref{mag.interaction}). As can also be seen in Fig.\,\ref{vapor}, the magnetic field is expanding with height, which causes the varying diameter of the ejection (cf. Fig.\,\ref{emiss} and movie provided in Appendix\,\ref{appendix}). Fig.\,\ref{vapor} also shows that the magnetic field is not helical.


\section{Onset of the plasma ejection}\label{onset}

In this section, the processes that trigger and drive the plasma ejection will be investigated. 
To study the onset of the plasma eruption in terms of time and space, peculiarities in the plasma properties must be looked for, mainly with regard to the current density, which is an indicator for plasma heating that can cause an eruption (Sect.\,\ref{formation1}).

When looking at the computational domain from the side, the ejection becomes clearly visible after it has emerged above the ambient transition region.
This is the case at a height of approximately 10.5\,Mm, between $t{=}39$\,min and 40\,min. The shape of the ejection appears tailed when ascending or descending and ellipsoidal when it is close to its maximum height. The ejection returns to the initial height at $t{=}51.5$\,min on the other side of the loop-shaped magnetic field lines. Subsequently, it slowly dissolves into the bright plasma and has completely disappeared after $t{=}56$ min.

To determine the trajectory of the ejection, the plasma that is part of the ejection must be defined. It proved to be useful to consider ejected plasma in the volume where the emissivity of  O\,{\sc{vi}} is observed to exceed $10^{-7}$ W m$^{-3}$. 
Thus, the surface of the ejection can be defined and its center of gravity localized. Tracing the center of gravity as a function of time allows us to determine the trajectory of the ejection. This procedure works best for the time periods after the ejection has detached from the ambient transition region, during its rising phase and before it touches the ambient transition region again when it falls back onto the solar surface.
The analysis of the plasma parameters along this trajectory is presented in Sect.\,\ref{formation2}.

\begin{figure*}
\includegraphics[width=\linewidth]{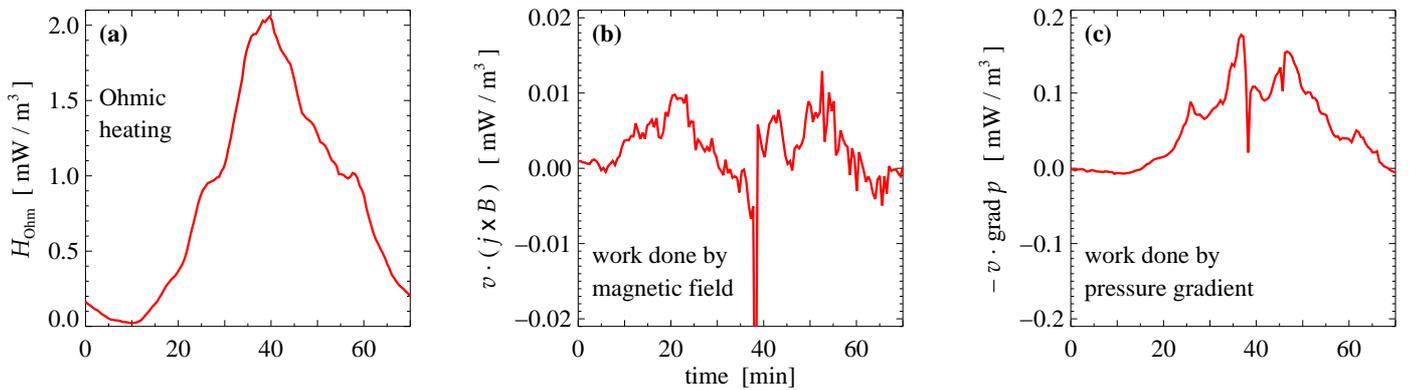}
\caption{Temporal evolution of plasma properties in the source region of the plasma ejection. The panels show (a) the Ohmic heating rate $H_{\rm{Ohm}}{=}\mu_0{\eta}j^2$, (b) the work done by the magnetic field and (c) the work done by the pressure gradient. Positive values in panels (b) and (c) imply that work is performed on the plasma, i.e., the kinetic energy of the plasma is enhanced.
All values represent averages in a box spanning 2\,Mm ${\times}$ 2\,Mm horizontally and 1\,Mm in height (cf.\ red box in right panels of Fig.\,\ref{emiss}).} 
\label{blob_start}
\end{figure*}

\subsection{Heating rate and pressure in the formation region}\label{formation1}

\begin{figure}
\begin{center}
\includegraphics[width=\columnwidth]{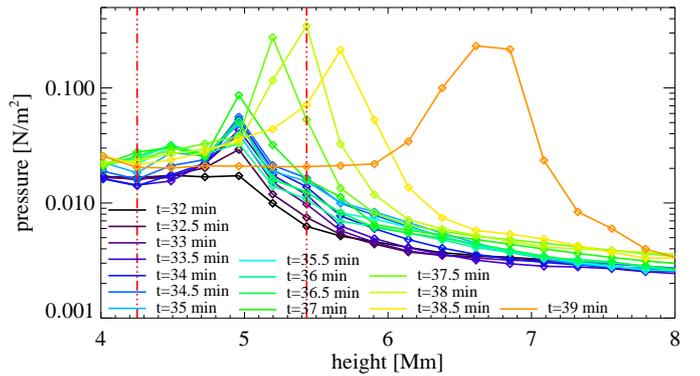}
\caption{Propagation of the pressure disturbance in the early phase of the ejection. During this interval, the ejection still rises almost vertically. The different curves show the (peak) pressure as a function of height in a vertical column indicated by the dotted lines in Fig.\,\ref{emiss} (bottom right panel). As time progresses, the location of the peak pressure moves upward.
The vertical extension of the red boxes in Fig.\,\ref{emiss} is indicated by the vertical dashed-dotted lines.
}  
\label{fig3}
\end{center}
\end{figure}

Fig.\,\ref{blob_start} visualizes the evolution of both the heating rate and the pressure in a small volume indicated by the red box in Fig.\,\ref{emiss} (the volume extension is identical for both the $x$ and the $y$ direction) chosen to capture the formation site of the ejection. Before the eruption, the Ohminc heating rate, $H_{\rm{Ohm}}{=}\mu_0{\eta}j^2$, continuously increases by approximately two orders of magnitude within 20 minutes (Fig.\,\ref{blob_start}, left) causing a temperature increase in the formation region. At the same time, the density increases through chromospheric evaporation. These occurances lead to a considerable rise in pressure (Fig.\,\ref{fig3}) that eventually results in the observed explosion at $t{\approx}39$\,min.

The rapid rise of the Ohmic dissipation, and thus the pressure, is limited to the small volume indicated by the red box in Fig.\,\ref{emiss} extending 2\,Mm${\times}$2\,Mm horizontally and 1\,Mm vertically). In general, the heating of the corona in the applied model is caused by the braiding of magnetic field lines \citep{Parker:1972,Parker:1988}. Numerical 3D MHD models for this process have been developed by \cite{Gudiksen+Nordlund:2002,Gudiksen+Nordlund:2005a,Gudiksen+Nordlund:2005b} for the first time. In these models, horizontal motions in the photosphere shuffle around the magnetic field which results in the braiding of magnetic field lines. Thus, currents are induced that are subsequently dissipated in the upper atmosphere. 

The photospheric driving by the horizontal motions in our 3D MHD model  \citep[see][]{bingert:2010} leads by chance, not by prescription, to a strong increase of the heating rate at the spot that will later become the origin of the plasma ejection.
This rapid increase of the pressure is comparable to an explosion that subsequently drives plasma away from the explosion site.

According to the realistic driver used to describe the photospheric motions, it appears plausible that a strong increase of the current density just above the chromosphere can also happen on the Sun, e.g., in the periphery of an active region or pore.

\subsection{Pressure along the trajectory of the ejection}\label{formation2}

The trajectory of the ejection must be defined before we investigate the ejection properties along the trajectory. We determined the volume of the ejected plasma as described above (i.e., where the emissivity of O\,{\sc{vi}} exceeds $10^{-7}$ W m$^{-3}$) 
and calculated its center of mass for each snapshot of the numerical model every 30\,s. 
Thus, the path of the ejection can be tracked from $t{=}40$\,min to $51.5$\,min, at which time the entire plasma ejection is visible and distinguishable from the background. For time steps prior to $t{=}40$\,min, the trajectory is obtained by a linear downward extrapolation (basically vertical).

For the early onset of the ejection before $t{=}39$\,min, the vertical column containing the formation site of the ejection was examined in detail just below the spot where the ejection becomes visible. This column has a quadratic cross section (2\,Mm\,$\times$\,2\,Mm) and is indicated in Fig. \ref{emiss} (bottom right) by the vertical dotted lines. The (peak) pressure in the vertical column is shown as a function of height for different time steps in Fig.\,\ref{fig3}. The dashed-dotted vertical lines in Fig.\,\ref{fig3} indicate the vertical extension of the red box shown in Fig\,\ref{emiss}. 

At $t{=}32.5$\,min, a small kink appears in the pressure profile at a height of approximately 5\,Mm, indicating a pressure anomaly. Remaining more or less at the same location, it develops into a local pressure peak within a few minutes. This marks the phase of temperature increase and evaporation of the plasma in response to the increased heating rate.
It is followed by the explosion, and, as a result, the pressure peak starts moving upward, as can be seen on snapshots at times $t{=}37$\,min to 39\,min. The part of the pressure disturbance moving downward is effectively dissipated by the dense plasma of the low atmospheric layers and is therefore not visible. 

Typically, a disturbance that travels upward into regions of lower density steepens and eventually forms a shock. However, dissipation can counteract this tendency to form a shock, and thus, the shape of the perturbation will be invariant on its way upward in the simulation box. This phenomenon of a soliton-like structure is indeed what is observed here. The pressure disturbance appears in the form of a soliton-like structure that keeps its amplitude and half width, leaves the top of the red box at $t{=}38.5$\,min and continues to travel upward.

\begin{figure}
\begin{center}
\includegraphics[width=\columnwidth]{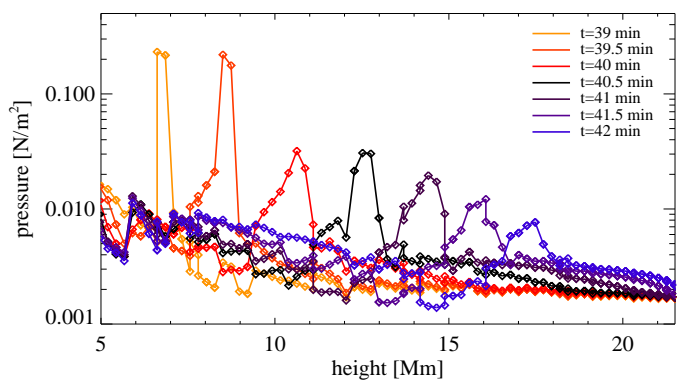}
\caption{Similar to Fig.\,\ref{fig3}, but for the peak pressure of the ejection along its trajectory (shown as a dashed line in the lower panels of Fig.\,\ref{emiss}, see also red line in Fig.\,\ref{fieldlines} for a 3D impression). For better comparison with Fig.\,\ref{fig3}, the pressure is plotted as a function of height, viz.\ altitude, along the trajectory (and not as a function of arc length along its path).
The first time step (39\,min, leftmost peak) corresponds to the last time step shown in Fig.\,\ref{fig3}.
}  
\label{fig4}
\end{center}
\end{figure}

The subsequent evolution of the pressure of the ejection on its way upward is shown in Fig.\,\ref{fig4}.  As the maximum pressure of the ejection is found roughly at its center of gravity, these values match the ones shown in Fig.\,\ref{fig3} (for better comparability,  the actual height in the simulation box is shown rather than the arc length along the trajectory). Thus, the lowermost location of the ejection as shown in Fig.\,\ref{fig4} at $t{=}38.5$\,min exactly corresponds to the uppermost location of the ejection shown in Fig.\,\ref{fig3}, which also occurs at $t{=}38.5$\,min. This shows that the plasma ejection continues to travel upward, still in a soliton-like fashion, but now the dispersion slowly becomes more important: The shape flattens as the ejection expands in the course of the flight.

\subsection{Energy considerations for the onset of the ejection}\label{energy}

In the previous section (\ref{formation2}) we outlined a scenario in which the increasing heating rate leads to an increase in pressure. The resulting pressure gradient drives the ejection --- similar to the situation of an explosion. We will confirm this scenario by investigating the energy budget.

\fig{fig3} shows that the pressure $p$ increases by 0.3\,N\,m$^{-2}$ within $\tau{=}5$\,min (the pressure starts increasing at $t{=}32.5$\,min and reaches its peak at about 37.5\,min). For an ideal monoatomic gas, the internal energy density $e$ is given through $e{=}3/2\,p$. Thus, the rate of change of the internal energy density is about $e/\tau=1.5$\,mW\,m$^{-3}$. This nicely corresponds to the Ohmic heating rate $H_{\rm{Ohm}}$ of 1.5 - 2\,mW\,m$^{-3}$ during that time (see \fig{blob_start}, left panel).
Thus, the pressure increase is indeed caused by the increased Ohmic heating rate of the plasma. The heating rate of some 2\,mW\,m$^{-3}$ is about a factor of 100 higher compared to normal regions at or just above the top of the chromosphere.

The acceleration during the early phase of the ejection can be derived from \fig{fig3}. At times $t=(38,38.5,39)$\,min the ejection reaches heights of about $z=(5.4,5.7,6.7)$\,Mm when traced by the peak pressure. 

From this, one can calculate the velocity to be about $v\approx25$\,km\,s$^{-1}$ and the acceleration to be $a\approx800$\,km\,s$^{-2}$. The density in this region is $\rho\approx5\cdot 10^{-12}$\,kg\,m$^{-3}$. Thus, the rate of change of the kinetic energy density of the ejection is about $\partial_t(\frac{1}{2}{\rho}v^2)={\rho}va\approx0.1$\,mW\,m$^{-3}$.  This number corresponds well to the amount of work done by the pressure gradient as shown in the right panel of \fig{blob_start} and gives a clear indication that the work done by the pressure gradient is pushing the plasma upward.

We can therefore conclude that the ejection is driven by the pressure gradient. The middle panel of \fig{blob_start} shows the work performed on the plasma by the magnetic field, or more precisely by the Lorentz force. This is about a factor of 10 weaker than the work performed by the pressure gradient, and it would not be strong enough to drive the ejection. Therefore, the Lorentz force does not play a significant role for the onset of the plasma ejection. When the ejection leaves the source region just before time $t=40$\,min, $v{\cdot}(j{\times}B)$ becomes negative, i.e., the plasma performs work on the magnetic field by pushing aside the magnetic field lines.

Finally, we note that the kinetic energy of the plasma ejection is about $10^{18}$\,J or $10^{25}$\,erg (assuming $10^9$\,kg for the average mass and 50\,km\,s$^{-1}$ for the average velocity of the ejection). This is about a factor of ten higher than the typical energy content of a nanoflare \citep[e.g.,][]{Parker:1988}.

\section{Dynamics and evolution of the plasma ejection}\label{evolution}

\subsection{Mass balance and chromospheric evaporation}\label{massbalance}

\begin{figure}
\begin{center}
\centerline{\includegraphics[width=0.55\columnwidth]{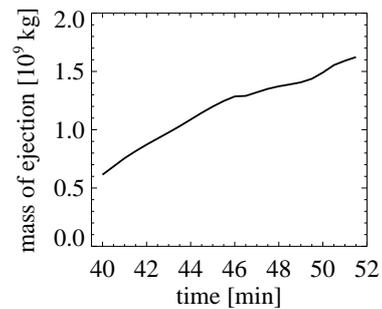}}
\caption{Temporal evolution of the mass of the plasma ejection.
\label{mass}
}
\end{center}
\end{figure}

After defining the volume of the plasma ejection, the total mass of the plasma ejection can be tracked. After a total of about $0.5{\cdot}10^9$\,kg has been ejected following the initial explosion,  the gas bubble acquires additional mass of about  $10^9$\,kg on its way along the magnetic field (see Fig.\,\ref{mass}). The total mass of the coronal part of the computational box, i.e., the mass of the plasma at temperatures above $10^6$\,K, is found to be about $5{\cdot}10^9$\,kg. Therefore, the cool plasma ejected up into the corona is a significant fraction of the total coronal mass. Still, it is only a tiny fraction of the entire mass in the computational domain owing to the high density of the photospheric and chromospheric plasma.

The (linear) increase of the mass of the ejection that flies through the corona as depicted in Fig.\,\ref{mass} cannot simply be accounted for by an accumulation of plasma in the corona before the lift-off of the ejection. Before and during the ejection, the (curved) tube along the trajectory of the ejection is filled by evaporated chromospheric material. The evaporation is caused by the increased heating rate before and during the plasma ejection. Thus, without the occurence of the eruptive ejection it is likely that a coronal loop (with increased density) would have formed through the long-known process of chromospheric evaporation following increased heating. After the explosion-type event, the ejection collects all material, which prevents the coronal loop from forming. The reason for this is that it sweeps away the evaporated material and transports it back to the chromosphere on the other side, where it is thermalized once the ejection hits the surface again.

\subsection{Dynamics}\label{dynamics}

\begin{figure}
\begin{center}
\includegraphics[width=\columnwidth]{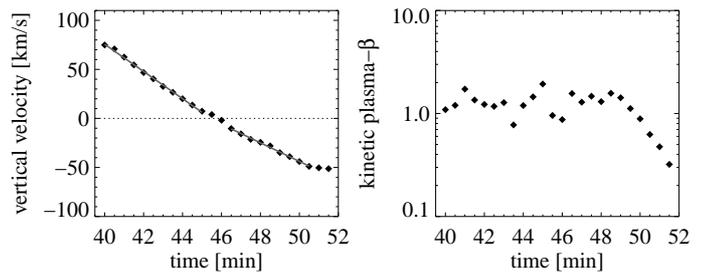}
\caption{{\it{Left:}} Temporal evolution of the vertical velocity; overplotted in gray are linear fits in the time intervals $t{=}40-45$\,min and $t{=}46.5-50.5$\,min. {\it{Right:}} Temporal evolution of kinetic plasma-$\beta$ at the center of gravity of the plasma ejection (see Eq.\,\ref{beta} in Sect.\,\ref{mag.interaction}).}  
\label{vel_beta}
\end{center}
\end{figure}

The vertical velocity $v_z$ of the plasma ejection is shown as a function of time in the left panel of Fig.\,\ref{vel_beta}. The dependency is characterized by the vertical component of the velocity of the center of gravity of the ejection. During the upward moving phase, the center of gravity is decelerated nearly constantly, until its initial speed of $v_z{\approx}75$\,km/s at $t{=}40$\,min has decreased to zero at $t{=}46$\,min, reaching the apex of the trajectory at a height of 21.5 Mm.
The ejection then continues (roughly) along the magnetic field lines and falls down with a roughly constant acceleration, reaching about 55\,km/s at $t{=}51$\,min, just before it hits the ambient transition region.

The deceleration and acceleration of the ejection are derived from a linear fit to the velocities in the time-periods $t{=}40$\,min to 45\,min and $t{=}46.5$\,min to 50.5\,min, respectively, during the rising and descending phases (Fig.\,\ref{vel_beta}, left panel, gray lines).
 The corresponding numbers are listed in Table \ref{accel}.

After the initial explosion-like acceleration, the ejection basically follows a ballistic flight. During the rising phase, the front of the ejection is stronger decelerated than the center of gravity (${\approx}255$\,m/s$^2$ for $t$=39.5-44 min, not shown in Fig.\,\ref{vel_beta}) because the ejection collects the material from the ambient corona at its front. Consequently, the shape of the ejection turns from tailed to ellipsoidal.

During the descending phase, the acceleration of the ejection is well below the solar surface gravitational acceleration (Table \ref{accel}). This is consistent with the mass balance as discussed in \sect{massbalance}. After the chromospheric evaporation, while the ejection collects the material along the trajectory, part of its momentum goes into the acceleration of the collected material, which is initially at rest. Consequently, the ejection is not accelerated as fast as a free falling body on its way down along the magnetic field lines.

\begin{table}
\begin{center}
\caption{Acceleration of plasma ejection (center of gravity).\label{accel}}
\begin{tabular}{ccc}
\hline
\hline
  phase [min]   & direction & acceleration [m/s$^2$]\\
\hline
  40.0 - 45.0   & upward   & 231\\
  46.5 - 50.5   & downward & 157\\
\hline
\end{tabular}
\end{center}
Note: Derived from linear fits to the vertical velocity shown in Fig.\,\ref{vel_beta}.
\end{table}

\subsection{Interaction with the magnetic field} \label{mag.interaction}

\begin{figure}
\begin{center}
\includegraphics[width=\columnwidth]{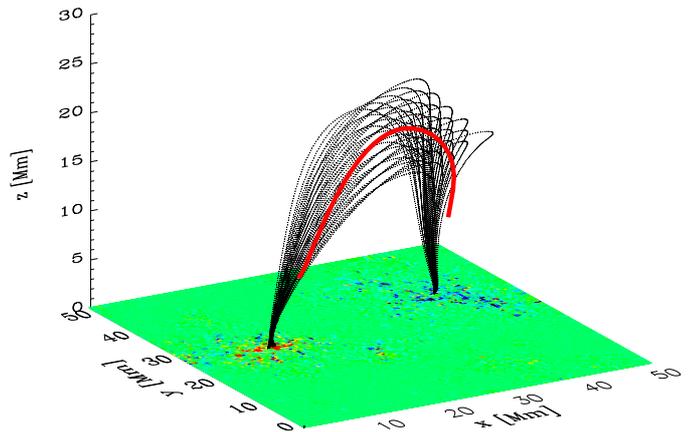}
\caption{Magnetic field configuration at timestep $t{=}15$\,min (well before the plasma ejection lifts off) together with trajectory of the plasma ejection (center-of-gravity) overplotted as a red line. The map at the bottom shows the vertical magnetic field at the bottom boundary of the simulation box scaled from $-$2000\,G (blue) to $+$2000\,G (red).}  
\label{fieldlines}
\end{center}
\end{figure}

The ejection roughly follows the magnetic field lines. However, because of its high momentum, it is capable to distort the coronal magnetic field (cf.\,Fig.\,\ref{fieldlines}). This is evident when investigating the kinetic plasma-$\beta$ term, which can be defined as the ratio of the kinetic energy density $\rho v^2/2$, and the magnetic energy density $B^2/2\mu_0$, i.e.,
\begin{equation}\label{beta}
\beta_{\rm{kin}} = \mu_0 \frac{\rho v^2}{B^2} ~ .
\end{equation}
Here, $\mu_0$ is the magnetic vacuum permeability, $\rho$ and $v$ are the mass density and velocity of the plasma ejection (at its center of gravity), and $B$ is the magnetic field strength.

In  Fig. \ref{vel_beta} (right panel) $\beta_{\rm{kin}}$ at the center of gravity of the ejection is shown as a function of time. Evidently, $\beta_{\rm{kin}}$ is approximately unity and can therefore displace the magnetic field, i.e., induce electric currents. These currents are visible as a tail behind the ejection in the current maps (right panels of Fig.\,\ref{emiss} and lower left panel in the movie attached to Fig.\,\ref{fig.movie}). This means that the plasma ejection bends the magnetic field.

Since the non-ideal MHD equations are solved in the numerical experiment, i.e., diffusivity is taken into account, the ejection will start moving through the magnetic field lines. The characteristic time for resistive diffusion is $\tau = \ell^2/\eta$, where $\ell$ is the length scale over which the ejection moves across the magnetic field, and $\eta$ is the magnetic diffusivity. Consequently, $\ell=(\eta\,\tau)^{1/2}$.
Thus, for a typical flight time $\tau=5$ min and a magnetic diffusivity $\eta = 10^{10}$ m$^2$/s used in the numerical experiment, our estimate implies that the trajectory of the plasma ejection can be shifted perpendicular to the direction of the magnetic field by $\ell \approx 1\dots2$ Mm. This number is consistent with the offset of the ejection trajectory perpendicular to the magnetic field lines (cf. Fig.\,\ref{fieldlines}).

\section{Visibility in AIA/SDO}\label{aia_section}

\begin{figure*}
\includegraphics[width=\textwidth]{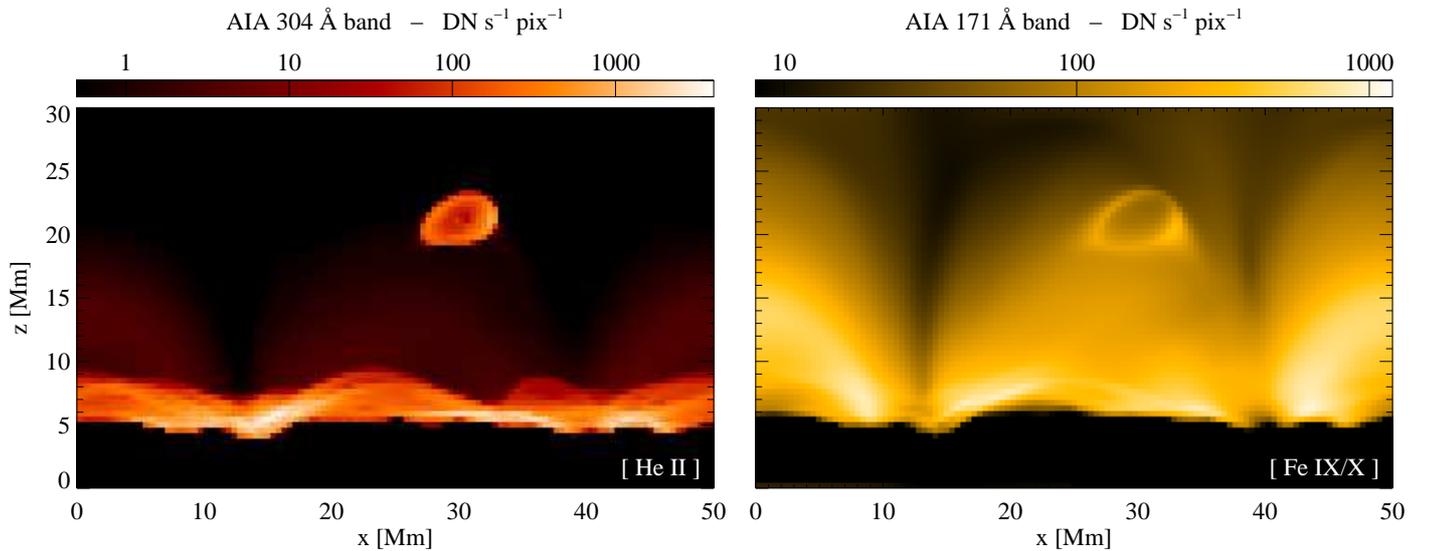}
\caption{Synthesized images as they would be observed at the limb by AIA/SDO in the 304\,\AA\ and 171\,\AA\ bands dominated by the He\,{\sc{ii}} line and the Fe\,{\sc{ix/x}} lines, respectively. 
These snapshots shows the ejection at $t{=}47$\,min, i.e., at the same time as the middle row of Fig.\,\ref{emiss} and as Fig.\,\ref{fig.movie}. The original synthesized images are downgraded to match the spatial resolution of AIA/SDO.
The intensity is given in expected AIA counts per pixel and second.}  
\label{aia}
\end{figure*}

As outlined in the introduction, cool structures are ubiquitously present in the hot corona. So far, we have focused on emission in O\,{\sc{vi}} and Mg\,{\sc{x}} to represent the hot and cool plasma. The Atmospheric Imaging Assembly \citep[AIA;][]{Lemen+al:2011} on the Solar Dynamics Observatory (SDO) is an appropriate instrument to observe ejections as those investigated in the present paper. The channels centered at 304\,\AA\ and 171\,\AA\ are dominated by lines of He\,{\sc{ii}} and Fe\,{\sc{ix/x}}, respectively. Their temperature contribution functions peak at ${\log}T\,[{\rm{K}}]=4.9$ and 5.9, i.e., not very far from the O\,{\sc{vi}}, and close to the Mg\,{\sc{x}} formation temperature (cf. Table \ref{lines}). Consequently,  the images synthesized for the AIA channels displayed in Fig.\,\ref{aia} do not show significant differences from those in O\,{\sc{vi}} and Mg\,{\sc{x}} presented in Figs.\,\ref{emiss} and \ref{fig.movie}.

To synthesize maps as AIA would observe the plasma ejection, we used the temperature response functions for the AIA channels as provided in SolarSoft.%
\footnote{http://www.lmsal.com/solarsoft/}
These are based on synthesized emission line spectra folded with the wavelength response of the filters. Applying the temperature response function at each grid point, multiplying by the density squared and then integrating along the line of sight provides the count rate to be expected in the respective band, i.e., the digital number DN/s/pixel. In Fig.\,\ref{aia} these maps are displayed for horizontal integration (along $y$) for the 304\,\AA\ band that shows the cool plasma at transition region temperatures and for the 171\,\AA\ band that displays the hotter plasma at coronal temperatures. The images are degraded to match the spatial sampling of AIA (0.6\,arcsec/pixel corresponding to 435\,km/pixel). The dynamic range of the AIA synthetic images in Fig.\,\ref{aia} is 1000 (304\,\AA) and 300 (171\,\AA), which is consistent with the estimates for the AIA dynamic range as provided in the AIA instrument paper \citep{Lemen+al:2011}. Also, the absolute values of the count rates of approximately 100\,DN/s/pixel are consistent with real solar observations of AIA.

The plasma ejection is hardly visible in the 171\,\AA\ band (right panel of Fig.\,\ref{aia}), it basically appears as a dark bubble with a bright ring. The reason for this is that the ejection is cool in the middle with the temperature rising to its circumference. However, we cannot really expect to see this ejection in the 171\,\AA\ channel in real observations. In the present case, the line-of-sight integration is over 50\,Mm only, i.e., the length of the computational box. On the real Sun, structures over a far greater length would contribute along the line of sight. Thus, the weak signature of the plasma ejection can be expected to be drowned in the background. If the density of the ejection is sufficiently high and if the temperature is sufficiently low, it might significantly absorb light at the wavelengths of the 171\,\AA\ channel through free-bound absorption \citep{anzer+heinzel:2005}. However, even then, ample structures in the foreground might contribute to the emission along the line of sight on the real Sun and cloud the visibility of the ejection in the 171\,\AA\ band and the other coronal channels of AIA.

The situation is different for the 304\,\AA\ channel. This band mainly displays cooler material at less than 100.000\,K and shows a clear signature of the plasma ejection during its flight. To our knowledge, no reports have been published so far on similar structures on ballistic orbits through the corona. However, it must be noted that SDO is fully operational since last summer only. We therefore propose to look for these structures in the AIA 304\,\AA\ data to confirm the counterparts of the (even cooler) structures seen in Ly-$\alpha$ by \cite{Schrijver:2001}, which were discussed in the introduction, and which are consistent with the plasma ejection described here. When analyzing the AIA 304\,\AA\ data one will have to struggle with the confusion of structures through the integration along the line of sight. Therefore, we cannot estimate yet if the mechanism proposed in this paper will be applicable for a wider range of processes seen in the 304\,\AA\ AIA channel.  However, we hope that the very high temporal resolution of AIA will allow us to see these cool ejecta flying in the background through the corona.


\section{Conclusions}\label{discussion}
We have analyzed a numerical experiment that describes the solar atmosphere above a small active region in which --- without external triggering --- a bubble of dense cool plasma is accelerated upward into the corona, moves along the magnetic field lines, and eventually falls back down onto the surface at a different place. The ejection is a direct consequence of the increased heating just above the chromosphere (and in the low transition region) for a short period of time. This can be considered an explosion, where magnetic energy is transformed effectively to thermal energy. As a consequence of the explosion, a pressure perturbation travels in a ballistic fashion away from the heating site along the magnetic field lines. 

While the heating rate in our model is based on magnetic processes (i.e., braiding of field lines and Ohmic dissipation of the induced currents), the transient event itself is basically a hydrodynamic phenomenon. The situation here is different from reconnection events, where plasmoids are driven away from the reconnection site by direct magnetic forcing. For the case described in this study, the increased heating leads to an explosion-type event, and the pressure gradient pushes the material up into the corona.

The increased heating rate along the magnetic field lines that eventually leads to the ejection also causes (a more gradual) evaporation of chromospheric material before and during the event. Thus, the bubble that flies through the corona and collects mass on its way contains plasma evaporated from both footpoints of its trajectory.

The ejection has sufficient kinetic energy to deform the magnetic field on its flight through the corona. This leads to current sheets in front of the ejection that propagate faster than the ejection and lead to heating (and evaporation) of plasma ahead of the ejection.

Because the ejection consists of cool plasma, it cannot be expected to be observed in coronal images of the Sun that display plasma at temperatures around and above $10^6$\,K, and which are available through the 171\,\AA\ channel of SDO/AIA. The ejection might be visible in data that image lower temperature plasma, such as the 304\,\AA\ AIA channel. However, more detailed studies will be needed to ascertain if these events are visible considering the confusion of structures along the line of sight.

The plasma ejection observed in our numerical experiment is consistent with the type (3) brightenings as seen in the TRACE Ly-$\alpha$ channel by \cite{Schrijver:2001} and discussed in the introduction. Even though the plasma seen in Ly-$\alpha$ is most probably cooler than the ejection described in this work, our proposal of an ejection driven by an explosion-type event might be a good explanation for structures flying on ballistic orbits through the corona. Further comparisons between numerical experiments and observations will have to show to what extent this is an ubiquitous mechanism for transient events seen through cool plasma in the hot corona.

{
\acknowledgements
The work of PZ and SB was supported by the Deutsche Forschungsgemeinschaft (DFG). We thank the anonymous referee for comments that helped improve the manuscript.
}


\clearpage

\begin{appendix}

\section{Temporal evolution of the plasma ejection} \label{appendix}

\begin{figure*}
\includegraphics[width=\textwidth]{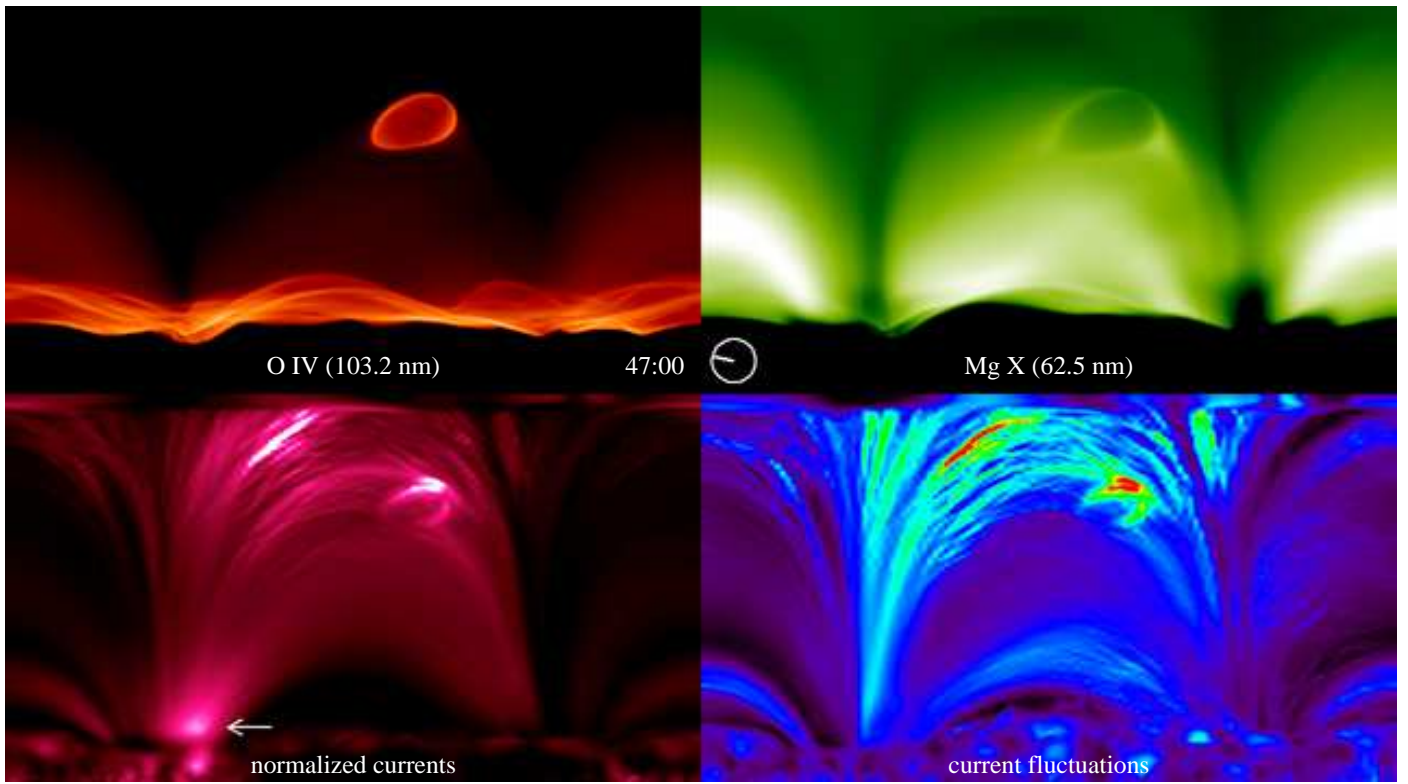}
\caption{%
Side view of the 3D computational box showing the synthesized emission in O\,{\sc{vi}} and Mg\,{\sc{x}} (top row), the normalized currents ${j^2_N}$ and the current fluctuations ${j^2_F}$ (bottom row; see Eqs.\,\ref{E:jnorm} and \ref{E:jfluct} and Appendix \ref{appendix} for details). Each panel covers 50\,Mm in the horizontal and 30\,Mm in the vertical direction. 
This snapshot shows the ejection at $t{=}47$\,min, i.e., at the same time as the middle row of Fig.\,\ref{emiss} and as Fig.\,\ref{aia}.
The arrow points at the location of the high energy input that causes the ejection.
A movie showing the temporal evolution over 1\,h is available in the online edition.
\newline
The movie is also available at http://www.mps.mpg.de/data/outgoing/peter/papers/plasma-ejection/blob.mpg (6\,MB).
\label{fig.movie}
}
\end{figure*}

The temporal evolution of the plasma ejection is best displayed by a movie showing the emission as well as the heating rate integrated through the computational box. To mimic the situation at the limb, a horizontal line of sight is chosen, e.g., the $y$-direction. The trajectory of the ejection is tilted by approximately 20$^\circ$ with respect to the $x$-direction (cf.\ Fig.\,\ref{fieldlines}), thus, these projections nicely show the ejection flying through the corona. 

The intensities $I(x,z;t) = \int_y \varepsilon(x,y,z;t) ~{\rm{d}}y$, are displayed in the top row of Fig.\,\ref{fig.movie}, where $\varepsilon$ is the emissivity at a given grid point $x,y,z$ and time $t$. 
To avoid Moir\'e patterns when displaying the images, the values for density and temperature were interpolated on a grid with higher spatial resolution before the emissivities were calculated using CHIANTI (see also Sect.\,\ref{phenomenology}). This procedure is similar to that of  \cite{peter+al:2004,peter+al:2006}. 
The intensities in Fig.\,\ref{fig.movie} are displayed on a logarithmic scale with dynamic ranges of 3000 (O\,{\sc{vi}}) and 50 (Mg\,{\sc{x}}) to achieve a better contrast.

The bottom panels in Fig.\,\ref{fig.movie} and the movie show the current $j$, a measure of the Ohmic heating rate (${\propto}j^2$), in two different normalizations. The lower left panel shows $j^2$ normalized by the average trend with height, i.e., the line-of-sight-integrated currents ($\int_y\dots{\rm{d}}y$) averaged along the $x$-direction and in time, $\langle\dots\rangle_{\displaystyle{x,t}}$,
\begin{equation}\label{E:jnorm}
{j^2_N} ~ (x,z;t) ~ = ~  \frac{\int_{y} j^2 (x,y,z;t)~{\rm{d}}y}%
       {\langle ~ \int_{y} j^2 (x,y,z;t)~{\rm{d}}y ~ \rangle_{\displaystyle x,t} } ~.
\end{equation}

This normalization is necessary to see structures in the heating rate (${\propto}j^2$), which when averaged horizontally, drops roughly exponentially with a scale height of about 5\,Mm \citep{bingert:2010}.

To emphasize the temporal variability, the heating rate was normalized at each grid point in the 3D computational domain by the temporal average, $\langle\dots\rangle_{\displaystyle{t}}$, at the respective grid point. This normalized quantity is then integrated along the line of sight, i.e.,
\begin{equation}\label{E:jfluct}
{j^2_F} ~ (x,z;t) ~ = ~  \int_{y} \frac{ j^2 (x,y,z;t) }%
                                           {\langle ~ j^2 (x,y,z;t) ~ \rangle_{\displaystyle t} } ~{\rm{d}}y ~.
\end{equation}
The term ${j^2_F}$  basically shows fluctuations along the line of sight, which is why it is indexed $F$. In the lower right panel of Fig.\,\ref{fig.movie}, this quantity is plotted using a rainbow color table, where violet represents low values of ${j^2_F}$ and red represents high values.

A potential magnetic field is used as an upper boundary condition {\itshape above} the computational domain in the 3D MHD model. Of course, the magnetic field inside the box is computed self-consistently based on the MHD equations. As a result, increased currents are observed near the top boundary. These currents have no impact on the magnetic structure and plasma heating in the main part of the box. They are small compared to those at lower heights, which are responsible for heating the coronal plasma (note that the average currents decrease exponentially with height).

\end{appendix}

\end{document}